\documentclass[aps,tightenlines,nofootinbib,preprint]{revtex4}
\usepackage{graphicx}
\newcommand{\be}{\begin{equation}}
\newcommand{\ee}{\end{equation}}
\newcommand{\bear}{\begin{eqnarray}}
\newcommand{\eear}{\end{eqnarray}}
\begin{document}

\title{Constraining renormalon effects
in lattice determination of heavy quark mass}

\author{Taekoon Lee}
\email{tlee@kunsan.ac.kr}
\affiliation{Department of Physics, Kunsan National University, 
Kunsan 573-701, Korea}

\begin{abstract}
The Borel summation technique of infrared renormalons is applied to
the lattice determination of heavy quark mass. With Borel summation
 a {\it physical} heavy quark pole mass and 
binding energy of
a heavy-light meson can be defined in a rigorous and
{\it calculable} manner.
A notable feature of the Borel summation, compared to the usual
perturbative cancellation of IR renormalons, is an automatic  scale
isolation. The two approaches of handling renormalon
divergence are compared in the B-meson as well as 
in an (imaginary) heavy-light meson with a mass much larger 
than the inverse of the
lattice spacing.

\end{abstract}

\pacs{}


\maketitle

The lattice simulation of the heavy quark effective theory of QCD
can provide potentially a most accurate
determination of the bottom quark mass. 
Essential in the calculation is constraining the unknown higher order
contributions in the perturbative matching of the lattice 
heavy quark effective theory and QCD. The matching relation for
the heavy-quark mass reads~~\cite{mar-sac}
\be
m_{\rm b}^{\rm pole} =M_{\rm B} - \bar\Lambda \,, 
\quad \bar\Lambda\equiv
{\cal E}(a) -\delta m(a) \,,
\label{rel1}
\ee
where $\bar\Lambda$ denotes the renormalized binding energy, 
$a$ is the lattice spacing, $M_{\rm B}$ is the B-meson mass,
$\delta m$ is the
mass shift in the static limit, and ${\cal E}$ is the binding energy
that is to be computed in lattice simulation. Here $1/m_{\rm b}$ 
corrections
are ignored. The mass shift as well as the matching relation between 
the pole mass 
and $\overline{\rm MS}$ mass can be computed 
in perturbation theory, and  with high order perturbative calculations
of the matching relations an accurate determination of
the $\overline{\rm MS}$ mass is possible from a precision
calculation of ${\cal E}(a)$. 

There is a caveat, however. As well known, the pole mass and 
mass shift suffer from renormalon divergences, and are not 
well-defined in perturbation theory. 
A well known approach for dealing 
this problem is to 
bypass the notion of 
pole mass (and any long-distance quantities) 
and  deal directly with a short-distance mass, imposing 
renormalon
cancellation perturbatively~~\cite{bb,bsuv,ms}.
This idea, when applied to
Eq.~~(\ref{rel1}), yields
\be
m_{\rm b}^{\overline{\rm MS}}=\Delta(a)
\left(1+\sum_{n=0}^\infty r_n(\mu/m_{\rm b}^{\overline{\rm MS}},\mu a)
\alpha_s(\mu)^{n+1}\right)
\label{pertmethod}
\ee
where $m_{\rm b}^{\overline{\rm MS}}$ is
the $\overline{\rm MS}$ mass satisfying
$m_{\rm b}^{\overline{\rm MS}} =
m_{\rm b}^{\overline{\rm MS}}(m_{\rm b}^{\overline{\rm MS}})$ and
$\mu$ denotes the renormalization scale and
$\Delta(a)\equiv M_{\rm B} -{\cal E}(a)$. The coefficients
$r_n$, which can be obtained from the perturbative expansions of
the pole mass and  mass shift, 
do not suffer from the renormalon divergence,
since the renormalon
divergence in the pole mass is cancelled by that in the mass shift.
Eq.~~(\ref{pertmethod}) truncated at next-leading order (NLO) was used
in Ref.~~\cite{ms99} in determining 
the $\overline{\rm MS}$ mass for the b-quark (A slightly
different approach may also be found in \cite{bali-pineda}).
For the sake of convenience, we shall call this method of handling
the renormalon divergence the perturbative cancellation method (PCM).

Let us now mention  two characteristics of the PCM.
We first notice that the coefficients
$r_n$ depend on two independent scales,
$m_{\rm b}^{\overline{\rm MS}}$  and $1/a$, 
through the renormalization scale. 
When the two scales are far separated, this can, in principle, delay
the 
convergence of the expansion~~(\ref{pertmethod}) and  be
problematic with low order perturbations.
This scale mixing is an 
unavoidable, generic problem 
as long as the renormalons in hard and soft quantities are 
cancelled in the perturbative way. Secondly, the PCM may not utilize
the known 
perturbative expansions of the involved quantities  to the 
fullest. Since $r_n$ is given by a linear combination of the
perturbative coefficients of the pole mass and mass shift to the order
$n$ the expansion~~(\ref{pertmethod}) can be determined only to the
order where both quantities are known.
For example, this prevented 
the PCM from using the next-next-leading order (NNLO)
calculation of the pole mass before the NNLO
calculation
of the mass shift.
Although
the  pole mass as well as mass shift are presently 
known to NNLO
a future next-next-next-leading order calculation of
one of these quantities cannot be utilized until both quantities
are calculated to the same order.
Our main point in this paper is that these undesirable  features
of the PCM
can be solved through the Borel summation of the divergent series
for the infrared-sensitive quantities.


A fundamentally different approach to the 
renormalon problem was proposed recently, in which the 
divergent perturbative expansions are
Borel summed to all orders~~\cite{surviving}. 
As well known, an infrared (IR) renormalon caused
large order behavior is of same sign and cannot be Borel summed.
This means that the Borel summation  
is ambiguous, depending on the integration contour 
of the Borel integral.

This ambiguity of Borel summation, however, should not appear in
physical observables, and this implies that the
ambiguities in the Borel
summation must cancel. In the case of the matching 
relation~~(\ref{rel1}),  the renormalons in the  pole mass and mass shift 
should cancel
since $M_{\rm B}$ must be free from the renormalon 
ambiguity.

When the perturbative series of the pole mass or mass shift
is Borel summed, with the positive real axis on the upper half-plane
as the contour of the Borel integral, the
ambiguity appears through the imaginary part of the Borel integral.
This ambiguity is known to be  proportional to $\Lambda_{\rm QCD}$ and 
independent of the  renormalization scheme and scale of the 
perturbation expansion~~\cite{beneke95}.
Applying  this Borel summation on the
perturbative series for the pole mass as well as the mass shift in 
Eq.~~(\ref{rel1}) we can now write 
the matching relation  as
\be m_{\rm b}^{\rm BR} =M_{\rm B} - \bar\Lambda^{\rm BR}\,,\quad
\bar\Lambda^{\rm BR}\equiv {\cal E}(a) -\delta m^{\rm BR}(a)\,,
\label{rel3}
\ee
where the `BR' quantities are defined as  the real parts of the
following
Borel integrals,
\bear
&&m_{\rm b}^{\rm BR}=m_{\rm b}^{\overline {\rm MS}}
\left(1+ \frac{1}{\beta_0}
{\rm Re}\int_{0+i\epsilon}^{\infty+i\epsilon}
e^{-b/\beta_0\alpha_s(\mu)}
\widetilde{ \cal M}_{\rm b}(\mu/m_{\rm b}^{\overline {\rm MS}},b)\, db
\right)\,, 
\nonumber\\
&&\delta m^{\rm BR}(a)=\frac{1}{a\beta_0}{\rm Re} 
\int_{0+i\epsilon}^{\infty+i\epsilon}e^{-b/\beta_0\alpha_s(\mu)}
\widetilde {\delta \cal M}(\mu a,b)\, db\,,
\label{rel4}
\eear
where ${\widetilde {\cal M}}_{\rm b}(\mu/m_{\rm b}^
{\overline{ \rm MS}},b)$ and 
$\widetilde{\delta {\cal M}}(\mu a,b)$  are the Borel transforms of 
($m_{\rm b}^{\rm pole}/m_{\rm b}^{\overline {\rm MS}}-1$)
and $a\delta m$,
respectively, and $\beta_0$,  the
one loop coefficient of the QCD $\beta-$function, is inserted here
for the normalization convenience of the Borel transforms.
The respective imaginary parts of the above Borel integrals are
identical, canceling each other on Eq.~~(\ref{rel1}),
and so 
can be ignored altogether, leaving only the real parts in the matching
relation. This means that, unlike the 
PCM where the renormalon cancellation
is implemented order by order perturbatively, the
renormalon cancellation in Borel summation is {\it exact}
from the beginning. The importance of this will be 
further discussed in the following.
Another nice feature of 
the Borel summation is that it preserves the original form of the 
matching relation, but unlike Eq.~~(\ref{rel1}), all involved 
quantities are now well-defined;
Eq.~~(\ref{rel1}) becomes rigorous through the Borel summation.

The exact nature of renormalon cancellation of the
Borel summation method has an important implication.
It resolves
the scale mixing problem of the PCM mentioned above and also 
allows one 
to utilize all the available perturbative expansions for the
pole mass as well as the
mass shift, since these two quantities are Borel
summed independently of each other.
Because $m_{\rm b}^{\rm BR}$ and 
$\delta m^{\rm BR}$ are  renormalization group (RG) invariant, 
one can choose
the renormalization scale, as well as even the scheme, 
for the pole mass 
independently of those for the mass shift.  This is in stark
contrast with the PCM where
the same renormalization scheme and scale should be chosen for both
the pole mass and mass shift to ensure renormalon cancellation.
This flexibility in choosing
the RG scheme and scale can be
exploited to optimize  
the Borel summations, which we will demonstrate later on
by choosing independent RG scales
in Borel summing the pole mass and mass shift.

The Borel summed $m_{\rm b}^{\rm BR}$ can, naturally, be called
a pole mass. By definition it has, when reexpanded in $\alpha_s(\mu)$,
exactly the same perturbative 
expansions as the (perturbative) pole mass, so is not a 
short-distance mass,
but nevertheless  is a well-defined quantity. This is also true
with $\delta m^{\rm BR}$, or $\bar\Lambda^{\rm BR}$.
Also, 
$m_{\rm b}^{\rm BR}$ and $\bar\Lambda^{\rm BR}$
are independent of the RG
scheme and scale as well as the lattice spacing $a$, so in this 
respect may be called `physical', although it does not mean that
the heavy-quark propagator, for instance, has a true pole at the BR mass.
They are also defined system-independently,
so once determined, can be used in any other systems.
As has been shown in~~\cite{surviving,lee03} and will be repeated
here later on, 
the ratio $m_{\rm b}^{\rm BR}/m_{\rm b}^{\overline {\rm MS}}$ 
can be calculated very accurately with the known perturbative
expansions of the pole mass. 
This means that the BR mass can be converted
accurately to the $\overline {\rm MS}$ mass, and vice versa, and 
can be freely used  wherever the use of a pole 
mass is more desirable. 



Defining a pole mass or a binding energy, formally, 
through the Borel summation is well-known, but 
the difficulty, however,
lies with the computability of thusly defined quantities;
If the Borel integrals of those quantities cannot be computed, 
to within the
necessary accuracy,
using the available information on the Borel transforms
the formal BR definitions
alone would be useless. 
The essential problem thus is how to rebuild the Borel transforms
sufficiently accurately, not only around the origin in the Borel
plane, where
the ordinary perturbation can be applied, but also about 
the renormalon singularity
that causes the renormalon divergence.
Clearly, because of the renormalon singularity, 
the usual perturbative expansions of the Borel transforms about the
origin alone would not suffice.

Since the renormalon ambiguity is $O(\Lambda_{\rm QCD})$,  
to solve the renormalon problem
it is necessary to 
calculate the BR quantities to an accuracy better than
$O(\Lambda_{\rm QCD})$. This requires an 
accurate description
of the Borel transforms in the region in the Borel
plane that contains
the origin as well as  the first IR renormalon singularity,
since the renormalon ambiguities arise
from the closest singularity to the origin.
With many orders of perturbative expansions, perhaps
many tens as may be inferred from the solvable instanton models in
quantum mechanics~~\cite{justin},
and the help from analytic continuations this could, in principle, be 
achieved, but is not possible in QCD since 
only the first few
perturbative terms are known. 
Without employing a very large number of perturbative expansions 
it is simply impossible to reconstruct
the renormalon singularity using the  perturbative Borel
transforms. It is important to realize that  
this calculability problem, not
the renormalon ambiguity, led to the abandonment of 
infrared-sensitive quantities
such as the pole mass.
We give in the following a brief account on how this 
calculability problem can be resolved by judiciously taking into
account the known properties of the renormalon singularity in addition
to the usual perturbative expansions.

In Refs.~~\cite{surviving,lee03} we have
shown that the above calculability problem can be resolved 
in the case of the pole mass and heavy quark potential
using an interpolation technique of the Borel transform, which we call
{\it bilocal expansion}.
A good description of 
the Borel transform about the origin can be obtained
from the first terms of perturbation, and the behavior of the
Borel transform about the
renormalon singularity
can also be learned, except for the residue of the singularity
that determines the overall normalization of the 
renormalon-caused large order behavior,
from the RG invariance of the renormalon ambiguity~~\cite{beneke95}.
Once the
residue is known, one can then 
interpolate the above two known behaviors
to obtain an accurate  description of the Borel transform
in the region of interest in the Borel plane. 
Of course, for this program to work
it is essential to calculate
the renormalon residue, and this problem
can be solved with the
perturbation scheme for the residue 
calculation developed in Refs.~~\cite{residue1,residue2}.
Fortunately, this  perturbative 
calculation of the residue for the pole mass  
turns out to converge very well, due to the overwhelming
dominance of the first IR renormalon in the $\overline{\rm MS}$ scheme,
and allows one to calculate the residue   to 
within a few percent of errors~~\cite{surviving,pineda0}.

The Borel transform in bilocal expansion has  two indices
that denote the orders of the expansions 
about the origin and the renormalon singularity.
For instance, in the case of the pole mass, the Borel transform
${\widetilde {\cal M}}_{\rm b}(\mu/m_{\rm b}^{\overline{ \rm MS}},b)$ 
is approximated by
${\widetilde {\cal M}}_{\rm b(M,N)}
(\mu/m_{\rm b}^{\overline{ \rm MS}},b)$
with the latter converging to the exact Borel transform in the
$M,N\to \infty$ limit. The indices $M$ and $N$ corresponds to
the order of 
perturbative expansions about the origin and about the renormalon
singularity, respectively.
With the perturbative calculations up to NNLO for the
pole mass and the four loop $\beta$-function, 
with the latter controlling
the expansion about the renormalon singularity,
${\widetilde {\cal M}}_{b(M,N)}$ can be obtained for
any combinations of $(M,N)$ with $M=0,1,2$ and $N=1,2$.
In the following we shall keep the bilocal expansions
to the highest known order, presently $N=1$, for the expansions about
the renormalon singularity, and consider expansions about the origin
with $M=0,1,2$, calling them the leading order (LO), NLO, and NNLO, 
respectively.
For the details of the bilocal expansion we refer the readers to
Refs.~~\cite{lee03,surviving}.

This bilocal expansion was shown to be very effective 
in Borel summing the pole mass as well as static interquark 
potential~~\cite{surviving}.
The Borel summed  pole mass and interquark potential
converge  rapidly under the bilocal expansion, and the 
resummed interquark
potential  agrees 
remarkably well with the lattice
potential up to the interquark distances
as large as about 0.7 ${\rm GeV}^{-1}$ (See \cite{pineda,sumino}
for the PCM approach to the static potential). 
Other applications in heavy quark physics may be found in
Refs.~~\cite{lee02,lee03,lee04,ag1,ag2}.

\begin{figure}
 \includegraphics[angle=0 , width=8cm ]{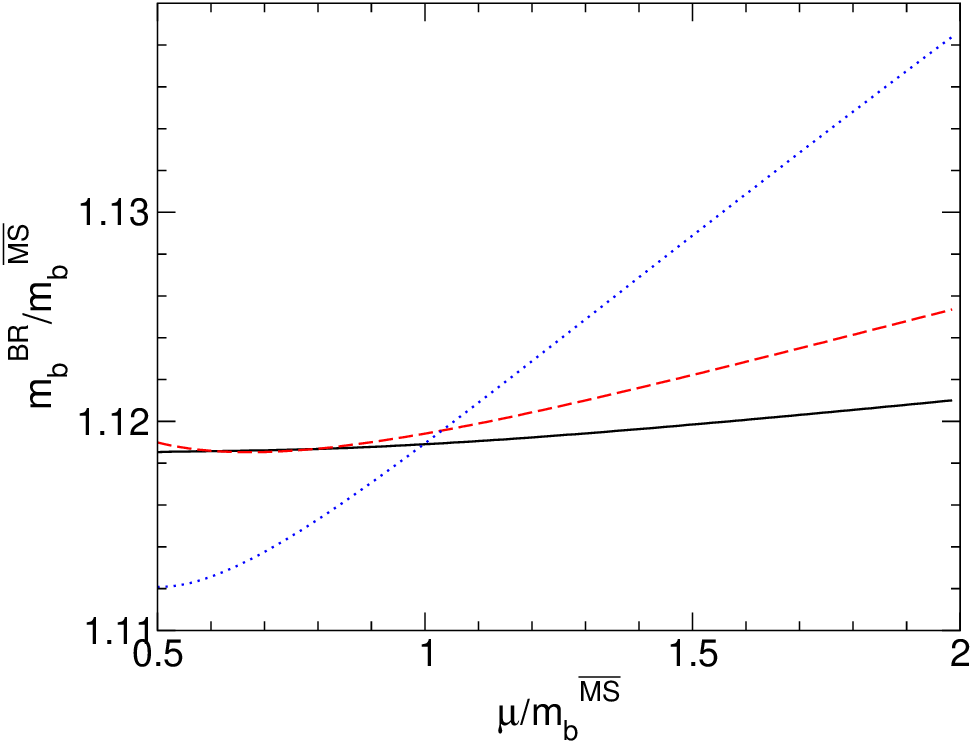}
\caption{$m_{\rm b}^{\rm BR}/m_{\rm b}^{\overline{\rm MS}}$ vs $\mu$.
The dotted, dashed, and solid lines are from using the Borel transforms
$\widetilde {\cal
M}_{\rm b (0,2)}, \widetilde {\cal
M}_{\rm b (1,2)}$, and 
$\widetilde {\cal M}_{\rm b (2,2)}$, respectively.}
\label{fig1}
\end{figure}

Now, going back to Eqs. (\ref{rel4})  we compute
the Borel summed pole mass
$m_{\rm b}^{\rm BR}$ as well as
the mass shift $\delta m^{\rm BR}(a)$ 
using the bilocal expansion described above. For the mass
shift we use the known first three terms of its perturbative expansion,
which allows us to compute the Borel transform 
$\widetilde{\delta {\cal M}}_{(M,N)}(\mu a,b)$ 
for $M=0,1,2$ and $N=1,2$. With these Borel transforms computing
the Borel integrals is then straightforward using a conformal
mapping. For the details of the
computation  we refer the readers to Ref.~~\cite{lee03} 
where the Borel summation
of the pole mass is described in detail.
The only difference with the present computations
is that the Borel integrals
in (\ref{rel4}) are with respect to $\alpha_s(\mu)$, whereas
in Ref.~~\cite{lee03}
it was with $\alpha_s(m_{\rm b}^{\overline {\rm MS}})$.
With this
change in the RG scale it is necessary to rescale 
the renormalon
residues accordingly. The RG invariance of the renormalon ambiguity
demands the renormalon residue scale linearly in $\mu$, which implies
the residue (of the first IR renormalon singularity) of  
${\widetilde {\cal M}}_{\rm b}(\mu/m_{\rm b}^{\overline{ \rm MS}},b)$
is given by $\bar C \mu/m_{\rm b}^{\overline {\rm MS}}$ where
$\bar C$ denotes the
reside at the scale $\mu=m_{\rm b}^{\overline{ \rm MS}}$, and the
residue of $\widetilde{\delta {\cal M}}(\mu a,b)$ is
$\bar C a\mu$.
For the quenched case ($N_f=0$) the
perturbative computation of the residue using the expansions of
the pole mass up to NNLO gives
\be
\bar C=0.424+0.168+0.035 =0.627\,,
\ee
where each terms denote the LO, NLO, and
NNLO contributions, respectively.

Throughout the paper we focus our attention only to
the quenched QCD, since
only the quenched lattice data for the binding energy are available,
and our main purpose is not to determine the $\overline {\rm MS}$ mass 
as precisely as possible but to make a comparison
of the two approaches for handling the renormalon divergence.
Hence in the following computations 
we also do not attempt to make detailed error estimates other
than those of perturbative origin.

The running coupling $\alpha_s(\mu)$ used in the following
computations is obtained using the four loop 
$\beta$-function and the lattice determination of $\Lambda_{\rm QCD}$
in quenched limit~~\cite{sommer}
\be
\Lambda_{\overline{\rm MS}}^{(0)}r_0=0.602\,,
\ee
where $r_0$ denotes the Sommer scale whose value is taken to be
$1/r_0=0.395$ GeV.

First,  we present the Borel summed pole mass in 
Fig. \ref{fig1} as functions of the RG scale.
Notice that the $m_{\rm b}^{\rm BR}$ at NNLO using the Borel transform
$\widetilde {\cal M}_{\rm b (2,2)}$ has a very small scale 
dependence, especially below $\mu=m_{\rm b}^{\overline{\rm MS}}$,
where the optimal scale is expected to be on.
Putting  $\mu=m_{\rm b}^{\overline{\rm MS}}$
and $m_{\rm b}^{\overline{\rm MS}}=4.31\,\, {\rm GeV}$, 
as an example,
we get
\be
\frac{m_{\rm b}^{\rm BR}}{m_{\rm b}^{\overline{\rm
MS}}}=1+0.11810+0.00044-0.00051
\label{ratio}
\ee
with each terms denoting the contributions from 
the Borel transforms
$\widetilde {\cal
M}_{\rm b (0,2)}, (\widetilde {\cal
M}_{\rm b (1,2)}-\widetilde {\cal
M}_{\rm b (0,2)})$, and 
$(\widetilde {\cal M}_{\rm b (2,2)}-\widetilde {\cal
M}_{\rm b (1,2)})$, respectively. Notice that 
the bulk of the Borel  summation is given by the LO contribution
and the higher order terms in bilocal expansion add only small
corrections. This indicates that the
leading order Borel transform, with the renormalon singularity
properly taken into account, already provides an excellent profile
of the true Borel transform in the domain of interest in the
Borel plane, with the
higher order terms contributing only small modifications .
The weak dependence on the RG scale
as well as the small size of the last two terms show that
the relation between the BR mass and $\overline{\rm MS}$ mass for 
the bottom quark can be precisely determined to within a few 
parts in $10^4$. 

\begin{figure}
 \includegraphics[angle=0 , width=8cm ]{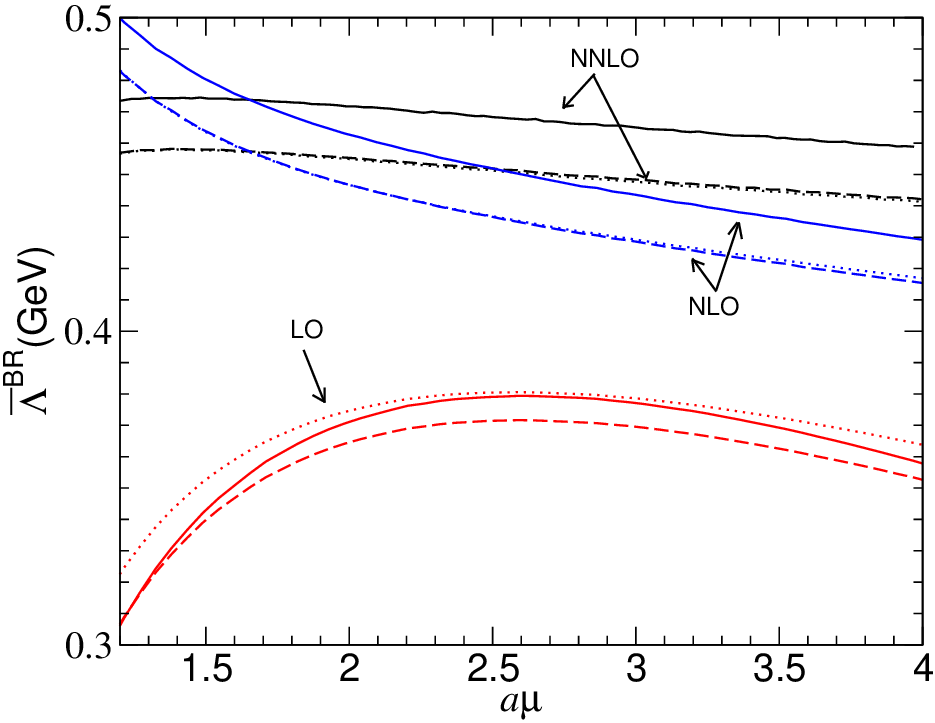}
\caption{The Borel summed binding energy vs $\mu$.
The dotted, dashed, and solid lines are at lattice spacings
$a^{-1}=2.12,2.91,$ and $3.85$, respectively, using
the leading order Borel transforms
$\widetilde {\delta \cal M}_{(0,2)}$,   
NLO Borel transforms 
$\widetilde {\delta \cal M}_{(1,2)}$, and
NNLO Borel transforms 
$\widetilde {\delta \cal M}_{(2,2)}$.
Notice that the lines for $a^{-1}=2.12$  and 2.91 
at NNLO (also at NLO) 
are almost on top of each other.}
\label{fig2}
\end{figure}

Now using the NLO and NNLO calculations for the mass shift 
\cite{ms99,scorz} we get the following bilocal Borel transform
for the mass shift to NNLO as
\bear
\widetilde {\delta \cal M}_{ (M,2)}(\xi,b)&=& 
\sum_{n=0}^M\frac{h_n(\xi)}{n!} \left(\frac{b}{\beta_0}\right)^n
+\frac{\bar C\xi}{(1-2b)^{1+\nu}}\left[ 
1 +\sum_{i=1}^2 c_i (1-2b)^i\right]\,,
\label{bilocalexpansion}
\eear
where
\bear
h_0&=&2.1173 -\bar C \xi (1+c_1+c_2) \,, \nonumber\\
h_1&=&3.7068 \log(\xi)-1.3053 -2 \bar C \xi
\beta_0[1-c_2+\nu(1+c_1+c_2)] \,,\nonumber\\
h_2&=& 6.4895 \log(\xi)^2-
1.8353 \log(\xi)+9.6538- \nonumber \\
     &&4 \bar C \xi 
     \beta_0^2[2+\nu(3+c_1-c_2)+\nu^2(1+c_1+c_2)]\,,
     \label{bilocal}
\eear
where $\xi=a\mu$ and 
\be
\beta_0= \frac{11}{4\pi}\,, \quad \nu=\frac{51}{121}\,, \quad
c_1=-0.2151\,, \quad c_2=0.1848 \,.
\ee

The  Borel summed binding energies vs $\mu$ from these Borel transforms
are shown in Fig. ~~\ref{fig2} 
 at lattice spacings $a^{-1}=2.12,2.91$, and $3.85$ GeV. 
Notice that $\mu$ is in units of the inverse lattice spacing that is
the only  scale relevant for the mass shift. This flexibility
of choosing the RG scale independently of that of 
the pole mass is a clear advantage over the PCM. 
The binding energies
${\cal E}(a)$ used in this calculation were taken from
the lattice data summarized in Ref.~~\cite{ms99} which read
\be
a{\cal E}(a)=0.61, \, 0.52,\, 0.46
\ee
at the lattice coupling $\beta=6.0, 6.2, 6.4$ that
correspond to $a^{-1}=2.12,2.91$, and $3.85$ GeV, respectively.
The lattice spacing from the lattice coupling was obtained 
following \cite{sommer}.

We first notice that the dependence on the RG scale
of the Borel summed 
binding energy $\bar\Lambda^{\rm BR}$ at NNLO
is less than $10\, {\rm MeV}$ over 
the  range of $\mu$ shown ($1.2/a\leq \mu\leq 4/a$), and,
remarkably, the binding energies at NLO and at NNLO for the 
two lattice spacings, $a^{-1}=2.12$ and 2.91, are virtually identical.
On the other hand the binding energies  
for $a^{-1}=3.85$ are about
15 MeV larger than those of the other lattice spacings, which
we suspect from the similarity of the line
shapes at NNLO (also at NLO) should be
largely due to the errors in ${\cal E}(a)$ from
lattice simulation  or in the relation between the
lattice coupling and lattice spacing than in perturbation theory.
We find the minimal sensitivity scales for the NNLO curves are at
$\mu^{*}\approx 1.41/a$ for all the 
three values of the lattice spacing, which turns out
to be close to the BLM scale that is about $1.45/a$~~\cite{ms99}.
Also the differences about the optimal scale
between the NLO and NNLO results are small.
All these indicate that the perturbative uncertainty in the
binding energy is well under control with the uncertainty
being at most $\pm 10 \,{\rm MeV}$.

Now reading the binding energies at the optimal scale we obtain
\be
\bar\Lambda^{\rm BR}= 0.458\,, 0.458\,, 0.474\, {\rm (GeV)}
\label{bindingenergy}
\ee
at $1/a=2.12\,,2.91\,,3.85$ (GeV), respectively. 
Using these values and the physical B-meson 
mass $M_{\rm B}\!=\!5.279 \,{\rm GeV}$ we get the corresponding
BR masses from 
Eq.~~(\ref{rel3})
\be
m_{\rm b}^{\rm BR}= 4.821\,, 4.821\,, 4.805\, {\rm (GeV)}\,
\label{brmass}
\ee
which have the same uncertainty as  
the binding energies in
(\ref{bindingenergy}),
ignoring the
negligibly small experimental uncertainty in $M_{\rm B}$.
Having the BR mass
 we can now 
obtain the  $\overline {\rm MS}$ mass  from
the relation between  the 
BR mass and $\overline{\rm MS}$, which is very
similar to (\ref{ratio}) but  
obtained using $\alpha_s(m_{\rm b}^{\overline {\rm MS}})$, with
$m_{\rm b}^{\overline {\rm MS}}$ determined self-consistently.
The result is summarized in Table ~~\ref{table1}.
The uncertainties in the extracted  $\overline {\rm MS}$ masses 
come   entirely
from the binding energies
$\bar\Lambda^{\rm BR}$, since the conversion
from the BR masses to  $\overline {\rm MS}$ masses adds only
a negligible error, about a few MeV. Thus
the perturbative uncertainty in the $\overline {\rm MS}$ 
mass is also estimated to be $\pm 10 \,{\rm MeV}$.

\begin{figure}
\includegraphics[angle=0 , width=8cm ]{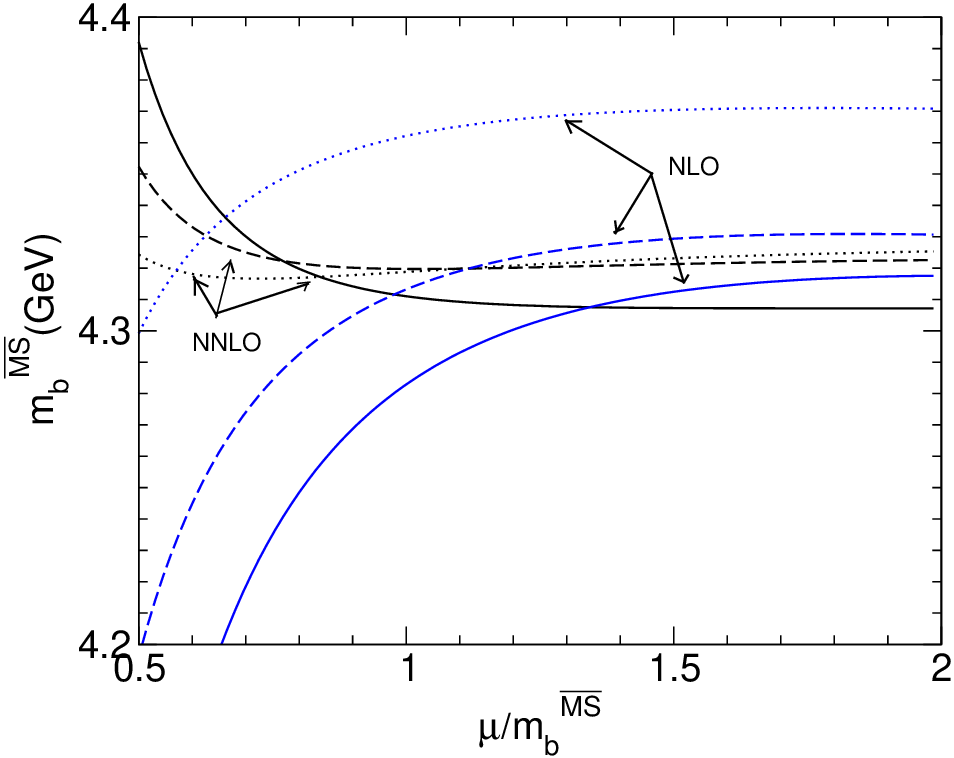}
\caption{The RG scale dependence of  $\overline {\rm MS}$ mass
determined in the PCM.
The dotted, dashed, and solid lines are at lattice spacings
$a^{-1}=2.12,2.91,$ and $3.85$, respectively.}
\label{fig3}
\end{figure}

Let us now see the scale dependence of the $\overline {\rm MS}$ mass
determined in the PCM based on Eq.~~(\ref{pertmethod}).
From the NNLO calculations
of the mass shift and  pole mass
the first three coefficients $r_0,r_1,r_2$ read
\bear
r_0&=& \frac{2.1173}{\Delta a} -0.4244 \,,\nonumber\\
r_1&=& \frac{1}{\Delta a}(3.7068 \log(\xi)-2.2039)-1.1820
-0.8488\beta_0\log(\chi) \,,\nonumber\\
r_2&=&\frac{1}{\Delta a} [6.4894 \log(\xi)^2-3.4084\log(\xi)+
7.7051-1.7972\beta_0\log(\chi)]-5.0606\nonumber\\
&&-1.6976\beta_0^2\log(\chi)^2-(0.8488 c_1+4.7279) \beta_0
\log(\chi)\,,
\eear
where $\chi=\mu/m_{\rm b}^{\overline{\rm MS}}$, and $\xi$ and $c_1$
are the same as in Eq. (\ref{bilocal}).

Fig. \ref{fig3} shows the NLO and NNLO results. At NLO,
the scale dependence appears to be much stronger than in the NLO
binding energy $\bar\Lambda^{\rm BR}$ in Fig. \ref{fig2}, 
and so is the dependence on the lattice spacing. A comparison
of the two figures shows that at this order
the perturbative uncertainty in
the  $\overline{\rm MS}$ mass from the PCM should be at least twice
as large as that of the BR method.

At NNLO the scale dependence as well as the dependence on the lattice
spacing are much improved, which are about less than
20 MeV for $0.7 \leq \chi \leq 2$.
For a comparison with the BR method
we  shall take the  $\overline {\rm MS}$ mass for each of the 
lattice spacings
at $\chi=1$, and the result is shown in Table \ref{table1}. 
\begin{table}
\par
\begin{center}
\begin{tabular}{l|ccc} \hline\hline
$a^{-1}$            &2.12    &2.91    & 3.85    \\ \hline
BR method           & 4.312 &4.312  &4.297   \\ 
PCM &4.319  &4.320  &4.311   \\ \hline\hline
\end{tabular}
\end{center}
\caption{\footnotesize The $\overline {\rm MS}$ mass 
$m_{\rm b}^{\overline {\rm MS}}$ determined in the Borel
summation method and PCM. The units are in GeV.}
\label{table1}
\end{table}

The  $\overline {\rm MS}$ masses at NNLO from the two approaches agree
remarkably well, the differences being smaller than 10 MeV.
This is a strong evidence that the renormalon cancellation is working
in both approaches. 

In contrast to the NLO results, at NNLO
the BR method does not seem to show a clear advantage over the PCM. 
The reason for this may be (i) at NNLO the PCM   
utilizes the NNLO calculation
of the pole mass as the BR method does and 
(ii) there is no large scale
separation in the B-system, the BLM scale of the
mass shift and the b-quark mass
being close. Between these two the second may be more
important: with almost no scale hierarchy in the system
the scale isolation feature of the BR method does not
have a 
room to show its advantage.

To confirm this we shall consider
an imaginary heavy-light
meson with its mass two orders of magnitude 
larger than the B-meson mass.
In the B-system the inverse of
the lattice spacing and the heavy quark mass were happen to be 
similar in size, so the effect of the scale mixing was a little 
subtle.
But in this new system with an exaggerated  scale difference
the usefulness of the 
scale isolation of the Borel summation will become
more obvious. Though the system we consider is not a real one 
this exercise could shed some insights on the scale mixing
problem in hierarchical systems, for example, like the 
top-pair threshold  production.

Let us call the heavy-light meson
H-meson and the heavy quark Q-quark,
and assume the H-meson mass is about two orders of magnitude 
larger than the B-meson mass, say,
\be
M_{\rm H}=500 \, {\rm GeV}\,.
\label{newmass}
\ee

Using  the $M_{\rm H}$ in (\ref{newmass}) and
the binding energies in (\ref{bindingenergy}), which
are independent of the heavy-quark mass,
we obtain from Eq.~~(\ref{rel3}) the BR mass
\be
m_{\rm Q}^{\rm BR}= 499.542\,, 499.542\,, 499.523\, {\rm (GeV)}\,
\label{Q-brmass}
\ee
at $1/a=$ 2.12, 2.91, 3.85 (GeV), respectively,
and from the following relation
between the BR mass and $\overline {\rm MS}$  mass, which was
computed similarly as in the b-quark system but 
with a rescaled strong coupling 
$\alpha_s(m_{\rm Q}^{\overline {\rm MS}})$,
\be
\frac{m_{\rm Q}^{\rm BR}}{m_{\rm Q}^{\overline{\rm
MS}}}=1+0.03638+0.00008-0.00004
\label{ratio1}
\ee
we obtain the corresponding $\overline {\rm MS}$ 
mass
\be
 m_{\rm Q}^{\overline {\rm MS}} =
 481.984\,,481.984\,,481.965 \,{\rm (GeV)}\,.
\label{msmass1}
\ee 
The mass ratio (\ref{ratio1})
was obtained self-consistently, 
using $\alpha_s(m_{\rm Q}^{\overline {\rm MS}})$ with
$m_{\rm Q}^{\overline {\rm MS}}$  obtained in 
(\ref{msmass1}); All three values of $m_{\rm Q}^{\overline {\rm MS}}$
give essentially the same ratio to within the accuracy.
Remarkably, the perturbative uncertainty in the extracted 
$\overline {\rm MS}$ masses 
in (\ref{msmass1}) should not be much larger than that of the
b-quark system from the BR method. This is because
the uncertainty in the relation~~(\ref{ratio1})
is small enough to add no new significant errors when the BR mass
is converted to the 
$\overline {\rm MS}$ mass.
If we assume
the uncertainty in the mass ratio (\ref{ratio1}) to be
$\pm 0.00004$, which is the size of the NNLO contribution,
then the conversion
from the BR mass to $\overline {\rm MS}$ mass causes an uncertainty of
only $\pm 20 \,{\rm MeV}$, and this is the dominant uncertainty
since the the error in $\bar\Lambda^{\rm BR}$ is the same as in
the B-meson. Thus the perturbative uncertainty in the 
$\overline {\rm MS}$
mass in (\ref{msmass1}) is estimated to be $\pm 20$ MeV.

\begin{figure}
\includegraphics[angle=0 , width=8cm ]{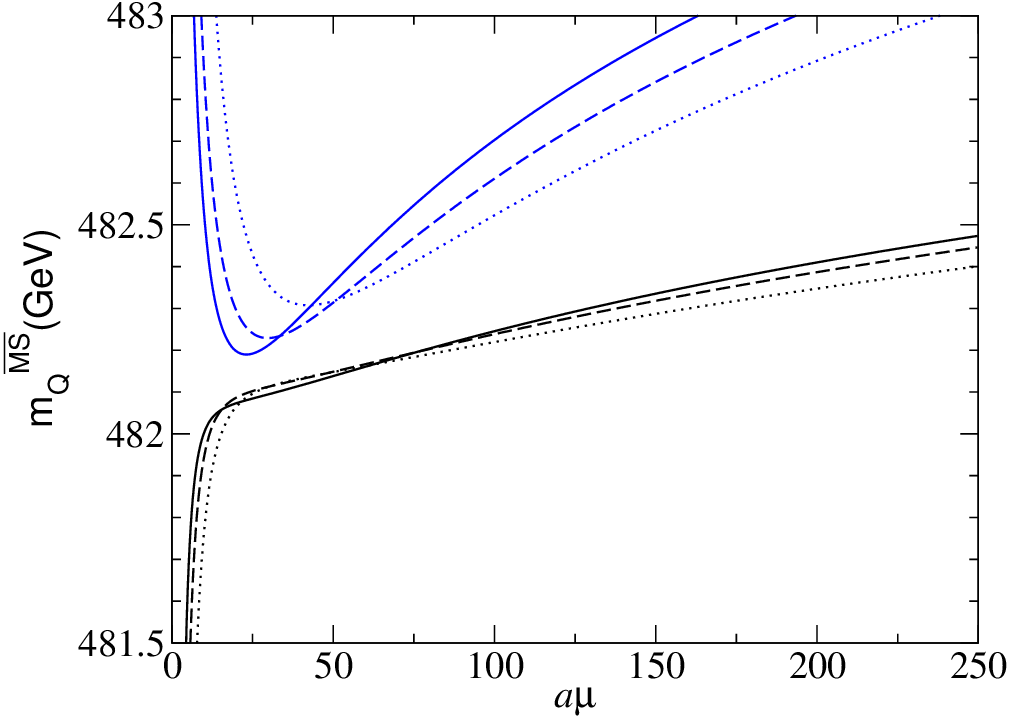}
\caption{The RG scale dependence of  $\overline {\rm MS}$ mass
determined in the PCM.
The dotted, dashed, and solid lines are at lattice spacings
$a^{-1}=2.12,2.91,$ and $3.85$, respectively. The upper three lines are
for the NLO and the lower three lines are for the NNLO results.}
\label{fig4}
\end{figure}

Let us now compare these $\overline {\rm MS}$ masses with those
obtained from using the PCM.
With (\ref{newmass}) and using Eq.~~(\ref{pertmethod})
truncated at $O(\alpha_s^3)$ with $m_{\rm b}^{\overline{\rm MS}}$
replaced by
$m_{\rm Q}^{\overline {\rm MS}}$,
we obtain the $\overline {\rm MS}$ 
masses in the PCM, which are plotted
in Fig. \ref{fig4} as functions of the RG  scale.
We first notice that the $\overline {\rm MS}$ masses at NLO have a
RG scale dependence of $O(1\,{\rm GeV})$ over the range
$1/a\leq \mu \leq m_{\rm Q}$, and the dependence on the
lattice spacing about a few hundred MeVs. The NNLO results
show improvement but still the scale dependence is about several
hundred MeVs while the dependence on the lattice spacing is rather
small, below 100 MeV, but still much larger than the uncertainty 
in the $\bar \Lambda^{\rm BR}$.
Thus the uncertainties
in the  $\overline {\rm MS}$ masses from the PCM
should be, at least, several hundred MeVs, much
larger than in the BR method. Even if one evaluates the
$\overline {\rm MS}$ mass at the optimal scale at which the
differences between the NLO and NNLO results become minimal
the obtained $\overline {\rm MS}$ masses
will be 
larger by more than 100 MeV than those of the BR method. 

Clearly, the strong dependence on the RG scale as well as on
the lattice spacing in the PCM comes from the scale mixing,
and this example shows that it can cause problems 
in systems with a large scale separation. The BR scheme 
solves this problem by isolating the heavy quark mass scale to
the BR mass and the soft scale, in this case the lattice spacing,
to the BR binding energy only,
resulting in a greatly improved heavy quark
mass determination. A good candidate to which the BR scheme
can be applied is the top threshold production, where
the cross section is computed 
from the Green's function obtained by solving a Schroedinger equation
involving a short-distance mass and 
a renormalon-subtracted heavy-quark potential, which again is
plagued by the scale mixing between the top quark mass
and the interquark distance~~\cite{topthreshold}. 
Our Borel summation can help this problem by summing the 
renormalons in the top pole mass and the interquark potential
independently, resulting in a Schroedinger equation
involving the BR top-quark mass  and interquark 
potential.
Here again the hard scale, the top mass, is confined to the BR 
mass and the soft scale, the interquark distance,
to the BR potential only, giving rise to
a complete scale isolation. With the PCM, on the other hand,
it may be necessary to
resum the $\log(v)$ terms  arising from the scale mixing~~\cite{hoang},
where $v$ denotes the heavy quark velocity, but with the BR scheme
this problem does not appear from the beginning.

To conclude, we have applied the Borel summation technique
based on the bilocal expansion of the Borel transforms, which
systematically takes into account the renormalon singularity as
well as the usual perturbative expansions, to the heavy quark
mass determination in lattice simulation.
The exact nature of the renormalon cancellation in the BR scheme
and the bilocal expansion
allow us to define a  calculable as well as physical
pole mass and  binding energy that are not short-distance
quantities, and give rise to a most useful feature of the BR
scheme, namely, the automatic isolation of the soft and hard
scales in the system. We have observed that this 
scale isolation results in a more consistent determination of the
heavy  quark mass than based on the PCM, and this becomes more visible 
as the system comes to have a bigger hierarchy in scale.
Although we have focused more on the practical advantages
of the BR scheme, we wish to emphasize that its more
important aspect is that it solves
the conceptual difficulty with the long-distance quantities, and
bring them to our avail.

\begin{acknowledgments}
I am very grateful to Tetuya Onogi for discussions from which this 
work was motivated. This work was supported in part by
Korea Research Foundation Grant (KRF-2004-015-C00095) and 
research funds from Kunsan National University.
\end{acknowledgments}

\bibliography{mb}

\end{document}